Demonstration of a white beam far-field neutron interferometer for spatially resolved small angle neutron scattering

Daniel S. Hussey[1], Houxun Miao[2], Guangcui Yuan[3], Dmitry A. Pushin[4,5], Dusan Sarenac[4,5], Michael G. Huber[1], David L. Jacobson[1], Jacob M. LaManna[1], and Han Wen[2]

1. Physical Measurement Laboratory, National Institute of Standards and Technology, 100 Bureau Dr., Mail Stop 8461, Gaithersburg, MD 20899, U.S.A.
2. National Heart, Blood, Lung Institute, National Institutes of Health, Building 10/B1D416, 10 Center Dr., Bethesda, MD 20892, U.S.A.
3. NIST Center for Neutron Research, National Institute of Standards and Technology, 100 Bureau Dr., Mail Stop 8461, Gaithersburg, MD 20899, U.S.A.
4. Institute for Quantum Computing, University of Waterloo, 200 University Ave. West, Waterloo, Ontario, N2L 3G1, Canada.
5. Department of Physics, University of Waterloo, 200 University Ave. West, Waterloo, Ontario, N2L 3G1, Canada.

**Abstract**

We provide the first demonstration that a neutron far-field interferometer can be employed to measure the microstructure of a sample. The interferometer is based on the moiré pattern of two phase modulating gratings which was previously realized in hard x-ray and visible light experiments. The autocorrelation length of this interferometer, and hence the microstructure length scale that is probed, is proportional to the grating spacing and the neutron wavelength, and can be varied over several orders of magnitude for one pair of gratings. We compare our measurements of the change in visibility from monodisperse samples with calculations which show reasonable agreement. The potential advantages of a far-field neutron interferometer include high fringe visibility in a polychromatic beam (over 30 %), no requirement for an absorbing grating to resolve the interference fringes, and the ability to measure the microstructure in the length scale range of 100 nm to 10 μm by varying either the grating spacing or neutron wavelength with a broad wavelength range and single set of gratings.

**Introduction**

Neutrons probe matter primarily via the strong nuclear force and thus provide a complimentary measure of the composition of a sample to electromagnetic probes such as x-rays. The interaction with the nucleus also provides a means for isotopic sensitivity, for instance allowing one to create contrast matched solutions composed of heavy and light water. In conventional attenuation based neutron radiography, one measures the local areal density (N t) from σ N t, where σ is the total scattering cross section with typical area of $10^{-24}$ cm$^2$ (that is 1 barn), N is the number density and *t* is the path length through the material. In neutron phase imaging, one measures a phase shift of the wavefront which is proportional to the local areal density from $b_c$ λ N t, where $b_c$ is the coherent scattering length (~$10^{-13}$ cm) and λ is the neutron wavelength (~$10^{-8}$ cm). Thus if the measurements have similar counting statistics and other uncertainties, neutron phase imaging would be ~1000 times more sensitive to variations in the areal density of an object. If there are fluctuations in the scattering length density (N $b_c$) due to a non-uniform microstructure, one observes small angle scattering (SAS). In interferometric phase imaging measurements, SAS results in a reduction of the visibility of the interference pattern.[1,2,3,4]

Since neutron sources are comparatively weak to x-ray sources, a challenge in developing neutron phase imaging is overcoming the low intensity that is an outcome of requiring small apertures to produce quasi-coherence in the transverse directions. The use of a neutron Talbot-Lau interferometer[5] was the first solution to this problem by requiring coherence in only one direction, enabling the use of a source grating which reduced the intensity by about 60 % rather than several orders of magnitude as was required for use of the transport of intensity method.[6] Neutron Talbot-Lau interferometers typically use a monochromatic beam ($\Delta\lambda/\lambda < 10$ %) and are composed of three gratings: a source grating of Gadolinium with period of order 1 mm; a phase modulating grating with combs etched in silicon with period of order 10 µm and aspect ratios of order 10; an analyzer grating composed of Gadolinium with period of order 5 µm. While one can create the source grating using laser etching or trench filling with Gadoxysulfide powder[7], the period and aspect ratio required for the analyzer grating remains a fabrication challenge. There is recent work on using Gd metal glasses which are able to reach high aspect ratio, small period Gd gratings that function well as analyzer gratings.[8] The use of a monochromatic beam is desirable as the Talbot distance where one obtains the self-image is inversely proportional to wavelength and the depth of the combs in the phase modulating grating is chosen to produce a $\pi$ phase modulation for the working wavelength. As well, the autocorrelation length of a first order neutron Talbot-Lau interferometer is somewhat constrained to a few micrometers, reducing the applicable range over which one can obtain quantitative small angle scattering information.[4]

Recent work on far-field interferometers for hard x-ray phase imaging has demonstrated that one can obtain phase images employing a polychromatic source and only phase modulating gratings.[9,10,11] These far-field interferometers are based on a moiré pattern formed by two (or more) phase modulating gratings, the general theory of which has recently been described.[12] The far-field interferometer has recently been demonstrated to work with neutron beams at both reactor and pulsed sources using polychromatic, monochromatic and pulsed time structures.[13] Briefly, the moiré pattern arises from an achromatic interference. Considering a pair of phase modulating gratings with the same grating period $P_g$, separated by a distance D, they produce an achromatic moiré pattern the period of which depends on the instrument length L as:

(1) $P_d = L P_g / D$,

So that the achromatic phase shift for a fringe displacement y is given $\delta\theta_0(y) = 2\pi y/P_d$. The visibility is determined by the parameters

(2) $\delta_1(\lambda) = \lambda L_1 D (L P_g^2)^{-1}$ and $\delta_2(\lambda) = \lambda L_2 D (L P_g^2)^{-1}$,

where $L_1$ is the distance of the source to the center of the first grating and $L_2$ is the distance between the second grating and the image plane. For $\pi/2$ phase gratings of square profile and 50% comb fraction, the fringe visibility is optimized at $\delta_1 = \delta_2 = \frac{1}{2}$. In terms of the $\delta$ parameters, the fringe period at the detector and the source period are then

(3) $P_d = \lambda L_2 (\delta_2(\lambda) P_g)^{-1}$ and $P_s = \lambda L_1 (\delta_1(\lambda) P_g)^{-1}$.

In order to create moiré fringes at the imaging plane for the two grating implementation, the source size must be limited to about $P_s/3$, though this can be composed of an array of slits. The wavelength dependence of the $\delta$ parameters influences the fringe visibility, which is a weighted average over the beam spectrum of the product of the ambiguity functions of the two gratings with the $\delta$'s as the autocorrelation distances (equation 16 in Ref. 12). Phase gradient and scatter images are obtained by placing the object either upstream or downstream of the gratings. The autocorrelation length probed by the interferometer is

(4) $\xi = \lambda z D / (P_g L)$,

where z is the distance of the sample to the detector if the sample is downstream of the gratings, or the distance to the source if the sample is upstream of the gratings. Taking $z = L/2$, $\xi = \lambda D/ (2 P_g)$, which corresponds to half the separation of the first diffraction order at the second grating. In this work, the grating separation was the primary means of varying $\xi$, with D ranging from about 3.5 mm to 40 mm.

**Experimental**

The experiments were conducted at the cold neutron imaging instrument on the neutron guide 6 (NG-6) at the NIST Center for Neutron Research (NCNR).[14] The instrument flight path enabled an overall length of L = 8.44 m, $L_2$ = 4. 24 m, and $L_1$ = 4.20 m. A vertical slit of 0.5 mm width masked the beam at the entrance; thus the transverse coherence length for a neutron wavelength of 0.5 nm was about 4 µm. Two silicon phase gratings of 2.4 µm period were used. Grating trenches were oriented vertically to avoid gravitational effects. The first grating was held fixed, while the second (downstream) grating could be manipulated in pitch, roll, yaw, and longitudinal separation. The mounting scheme was such that the closest approach of the two gratings was about 3 mm. Both gratings were then mounted on a rotary table to adjust the overall assembly yaw and height adjustment to toggle the position between the monochromatic and polychromatic beams (15 cm displacement). The data in Figure 1 was obtained with an amorphous silicon detector in direct contact with a 300 µm thick LiF:ZnS scintillator screen. The spatial resolution of this system is about 250 µm, and 100 exposures of 1 s were averaged. The data in Figure 2 were obtained with an Andor sCMOS NEO camera viewing a 150 µm thick LiF:ZnS scintillator with a Nikon 85 mm lens with a PK12 extension tube for a reproduction ratio of about 3.7, yielding a spatial resolution of about 150 µm.[15] To reduce noise in the sCMOS system, the median of three images with 20 s exposure time were used for analysis.

The wavelength distribution of the imaging instrument at NG-6 has not yet been measured. The simulated wavelength distribution can be approximated by a Maxwell-Boltzmann distribution with temperature 40 K giving a peak neutron wavelength of about 0.5 nm. While the far-field interferometer is achromatic, the visibility is maximized for wavelengths which experience a π/2 phase shift. To provide a π/2-phase shift for neutrons with a 0.5 nm wavelength, the grating depth in silicon should be 15.1 µm. However, there were 5 available gratings for the experiment, of varying depths, all with period of 2.4 µm and etched with a Bosch process. The gratings were fabricated at the NIST Nanofabrication facility.[16] We measured the visibility obtainable with three combinations of G1 and G2 after aligning the relative roll between G1 and G2 and scanned their relative separation. Since the gratings employed in this experiment were not optimized for this spectrum, it is possible that higher visibility could be obtained.

The raw images were converted to attenuation, phase gradient, and fringe visibility images using a Fourier transform method that requires a reference image with no sample and a single structured image of the sample. The full details of the algorithm can be found elsewhere.[17] As a note, without phase stepping data, the algorithm demodulates the images so that the pixel pitch of the reconstructed images is that of the fringe spacing; with phase stepping data, the reconstructed data have the same pixel pitch as the input data. In this work, only the attenuation and visibility images were analyzed. The attenuation image is the log-transform of the ratio of the mean of the fringes, $-\ln(H0_s / H0_r)$; the scattering image is the reduction in fringe visibility exceeding attenuation, is given as $\mu_d\, t = -[\ln(H1_s / H1_r) - \ln(H0_s / H0_r)]$; in both cases the subscript s (r) refers to the sample (reference) data, t is the sample thickness.

The equivalent inverse length scale probed within the sample (i.e. the Q-vector) is $q \sim 2\pi\, P_g\, (D\lambda)^{-1}$. By scanning either (or both) the inter-grating spacing or the wavelength one obtains a measure of the pair-wise auto-correlation function $G(\xi)$. For the geometry of the present setup, the q-ranges from about $10^{-2}$ nm$^{-1}$ to $10^{-4}$ nm$^{-1}$. It would be feasible to approach higher q-ranges with either larger period gratings or grating mounting that enabled a closer approach. To reach higher q-ranges, one could use finer pitched gratings or a detector with higher spatial resolution may enable resolving the fringes for large separations, but the fringe visibility maybe reduced to an impractical level. That said, the q-range of a typical first order Talbot-Lau neutron interferometer is more tightly restricted due to the constraints from the wave-length specific Talbot-distance and the proximity of the sample with the gratings.

**Results**

The pair of gratings that produced the highest visibility were installed and finely aligned with regards to pitch, roll and yaw, and overall yaw alignment with the beam. A scan of the change in visibility with respect to the inter-grating spacing, D, was conducted, shown in Figure 1. Visibility was defined as 2*H1/H0 or equivalently (Imax-Imin)/(Imax+Imin). A distribution of the visibility of the far-field interferometer flat field was observed, as can be seen in the raw images and persists for the range scanned D. The visibility distribution may be connected to the wavelength distribution of the flat field or to inhomogeneities in the etch depth of the gratings. A ray-tracing simulation of the beam line shows that the average wavelength in the beam center (corresponding to the right hand side of the images) is about 0.55 nm and increases to about 0.8 nm where the highest visibility is observed on the left.

To assess the quantitative measurement of the microstructure with the far-field interferometer, a monodisperse solution of (1.97 ±0.34) µm diameter polystyrene spheres with sample thickness 1 mm and 4 mm were imaged. The polystyrene spheres were in a solvent mixture of 50 % H2O, 50 % D2O (by volume) to give a sphere volume fraction $\phi_v$=0.05 so that the mass densities of solid spheres and solvent were nearly matched to prevent sedimentation during the measurement. The fringe visibility reduction was measured over the 1 cm diameter sample area and is shown in Figure 2. As the coherence length is increased, the fringe visibility reduction for dilute spheres of radius r should asymptotically approach $\mu_d = 3/2\, \phi_v\, \Delta\rho^2\, \lambda^2\, r$. The data show a clear decrease in $\mu_d$ with grating separation (and hence ξ). We believe this is due to the fact that as the grating separation is increased, the wavelengths that contribute most strongly to the visibility are shifted to shorter wavelengths, see Equation (2). Working on this assumption, we model the change in the average of the wavelength distribution as a ramp function

(5) $\lambda(D) = \Theta(D_1 - D)\, \lambda_1 + \Theta(D - D_2)\lambda_2 + \Theta(D-D_1)\Theta(D-D_2)\, \{\lambda_1 + (\lambda_2 - \lambda_1)(D - D_1)/(D_2 - D_1)\}$.

The four free parameters were determined from a nonlinear least squares fit of the data using the pair correlation function for dilute spheres, which is shown as the solid and dashed lines in Figure 2b. From the fit, $\lambda_1$ = (0.643±0.013) nm, while $\lambda_2$ = (0.515±0.003) nm, which are reasonable values for the cold neutron spectrum of the NG6 imaging beamline. While the model is somewhat ad hoc, it is a reasonable fit to the data. This situation would not present if the measurements were made with a monochromatic beam, either using a double crystal monochromator, velocity selector, or time of flight measurements.

Using equation 4 and 5 to calculate $\xi$, Figure 2c shows $\mu_d$ for both the 1 mm and 4 mm thick samples. The predicted $\mu_d$ from a dilute monodisperse solution of polystyrene spheres is also plotted. The data show clear oscillations, which may indicate that multiple scattering is present or that the spheres are sticking together to create twin dumbbells. In the latter case the auto-correlation (1-G($\xi$)) has an expected oscillation period of (2/π)*diameter = 1.25 µm, which approximately matches the observed period of about 1.5 µm. That the oscillations are more prevalent in the 1 mm sample than the 4 mm sample is surprising, but the 1 mm sample showed a faster settling rate than the 4 mm thick sample, as shown in Figure 3, indicating that there was likely a higher number of particles that formed dumbbell pairs.

**Conclusions**

We have shown that using only phase modulating gratings, one can create a neutron far-field interferometer which creates neutron small-angle scattering images. One can easily tune the autocorrelation length probed by this far-field interferometer to obtain a measure of the microstructure of an object, thus providing multi-scale imaging capability. After correcting for a change in wavelength, the measured reduction in fringe visibility agree reasonably well with that expected from a monodisperse solution of spheres. The fairly broad range of probed autocorrelation length also allowed observation of a likely short-range ordering of the spheres. Finally, since the far-field interferometer has a strong achromatic character, the method is ideally suited for use at a pulsed neutron source, enabling more quantitative measure of the visibility and phase gradients. This, combined with the removal of absorption gratings, is a significant improvement over the Talbot-Lau interferometer which is designed for a narrower wavelength band.

**Acknowledgments**

The authors wish to thank E. Baltic for technical assistance on the beamline, and useful discussions with M. Bleuel and J. Barker. This work was supported by the U.S. Department of Commerce, the NIST Radiation Physics Division, the Director's office of NIST, and the NIST Center for Neutron Research. This work is also supported by the Natural Sciences and Engineering Research Council of Canada (NSERC) Discovery program, and by the Collaborative Research and Training Experience (CREATE) program.

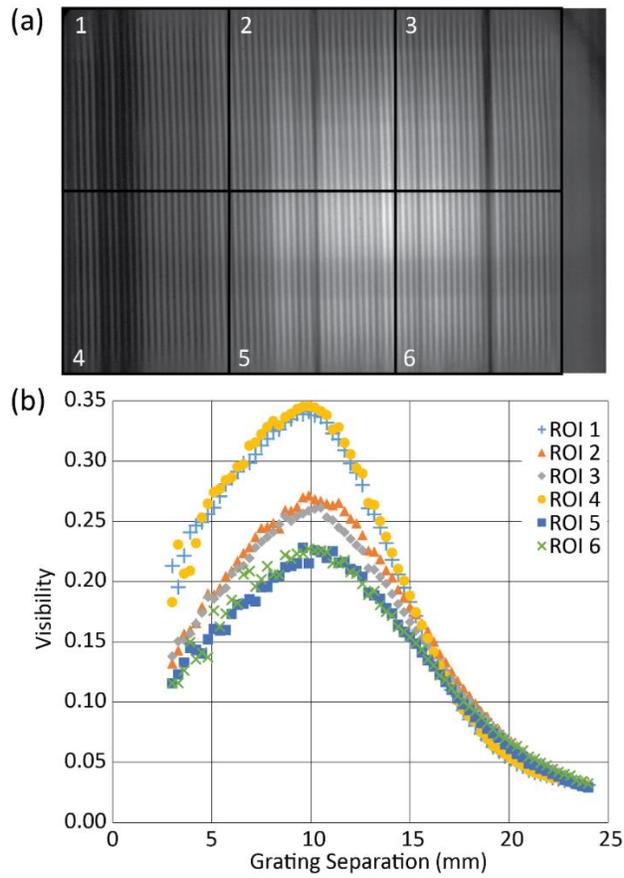

Figure 1. (a) The raw moiré pattern for 10 mm grating separation. (b) The visibility of the moiré pattern as a function of grating separation for the 6 regions of interest indicated in (a). The spatial resolution of the detector was 0.25 mm and limited the range of the visibility measurements in this case.

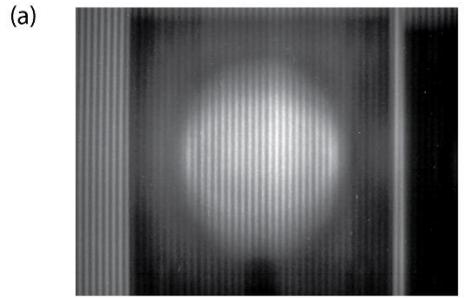
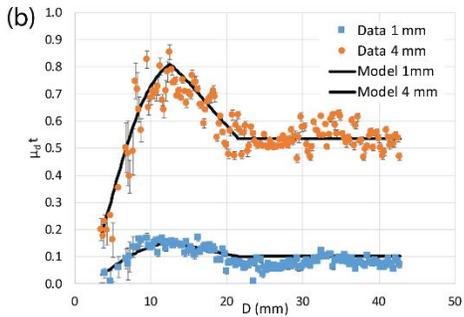
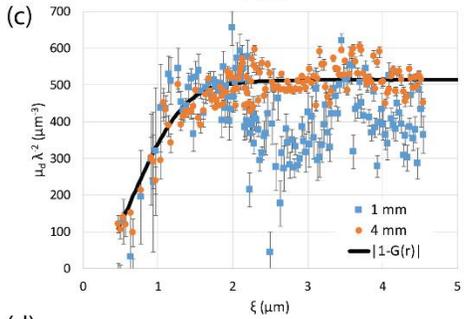
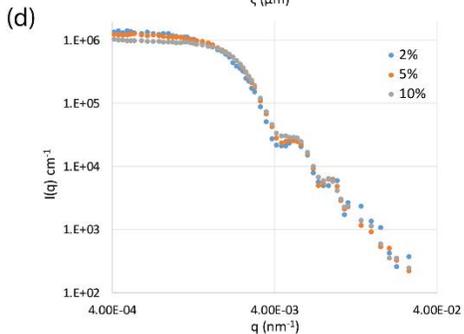

Figure 2 (a) Raw image of the 1 mm thick polystyrene sample at a grating separation of 13.5 mm.  (b) Measured visibility reduction as a function of grating spacing.  The uncertainty bars are derived from the one standard deviation of measured fringe visibility reduction over the imaged area of the cuvette.  As such, the actual uncertainty is likely larger due to other systematic measurement uncertainties that have yet to be fully quantified.  The solid lines are a model assuming a dilute mixture of spheres, and the mean of the wavelength distribution changes linearly between two separations, which were determined from a non-linear least squares fit.  (c) Using the lambda ramp function, $\mu_d \lambda^{-2}$ is plotted vs calculated pair-wise auto-correlation $|1-G(\xi)|$ for a dilute suspension of spheres.  The data show an oscillation, indicating that clustering (short-range order) of the spheres or multiple scattering is being observed.  This might also indicate that the correction for the change in wavelength needs to be improved, measurements with a monochromatic beam would not suffer from this systematic effect.  (d) USANS measurements of three volume fractions of spheres, from which the diameter was measured; the oscillations are due to the monodispersity of the spheres.

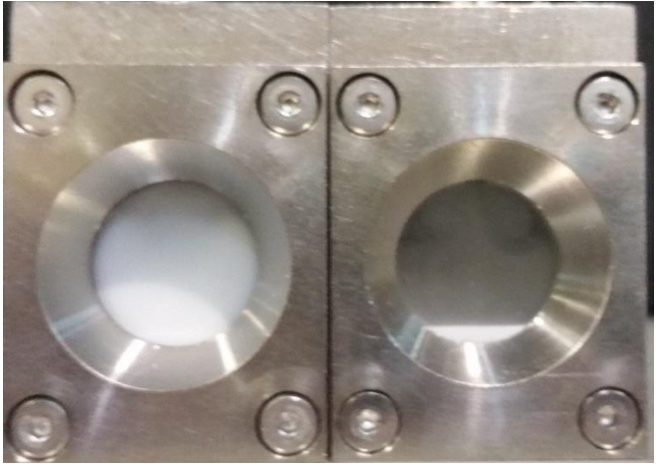

Figure 3 Photo of the sample holders for the polystyrene sphere solution about 3 weeks after the measurements.  The 4 mm thick holder is on the left and is clearly turbid, while the 1 mm thick holder is on the right showing that the spheres have settled out.